\RequirePackage[2020-02-02]{latexrelease}
\documentclass[aip,twocolumn,letterpaper]{revtex4}

\usepackage{fullpage}
\usepackage[margin=1.0in]{geometry}
\usepackage{graphicx}  
\usepackage{graphics}
\usepackage{appendix}  
\usepackage{xcolor}
\usepackage{amsmath}
\usepackage{hyperref}
\usepackage{subfig}

\def\h#1{\hat{#1}}

\def\avg#1{\langle{#1}\rangle}

\def\mb#1{\mathbf{#1}}
\makeatletter
\DeclareRobustCommand\ref{%
	\@ifstar\@refstar\T@ref
}%
\DeclareRobustCommand\pageref{%
	\@ifstar\@pagerefstar\T@pageref
}%
\makeatother

\begin{document}
\title{Current potential patches}
\author{Todd Elder\textsuperscript{1,2,*} and Allen H Boozer\textsuperscript{1}}
\affiliation{\textsuperscript{1} Columbia University, New York, NY, 10027}
\affiliation{\textsuperscript{2} University of Maryland, College Park, MD, 20737}
\affiliation{\textsuperscript{*}Corresponding author: Todd Elder, \textit{tme2123@columbia.edu}}

\begin{abstract}
	A novel form of the current potential, a mathematical tool for the design of stellarators and stellarator coils, is developed. Specifically, these are current potentials with a finite-element basis, called \textit{current potential patches}. Current potential patches leverage the relationship between distributions of magnetic dipoles and current potentials to explore limits of the access properties of stellarator coil sets. An example calculation is shown using the Helically Symmetric Experiment (HSX) equilibrium, demonstrating the method's use in coil design and understanding the limits of the access properties of coil sets. Current potential patches have additional desirable properties such as of promoting sparse current sheet solutions and identifying crucial locations of shaping current placement. A result is found for the HSX equilibrium that shaping currents covering only 22\% of the winding surface is sufficient to produce the equilibrium to a good accuracy, provided a toroidal field is generated by an exterior coil set.
\end{abstract}

\date{\today} 
\maketitle

\section{Introduction}
Stellarators are devices for magnetic confinement fusion (MCF) which have the potential to provide nearly limitless clean energy. Modern stellarators are optimized a during their design using supercomputers and novel calculational methods, yielding steady-state MCF devices which mitigate physical problems common to both stellarators and tokamaks such as magnetic island formation and energetic particle loss. As such, stellarators are at the cutting edge of MCF technology.  The experimental success of the optimized W7-X stellarator in Germany has confirmed that excellent plasma properties can be achieved by computational design \cite{ConfirmW7XToplogy:2016}.

A major barrier in the development of the stellarator is coil complexity. Stellarator coils suffer from poor access properties and tight tolerances. In order for the stellarator to become a device as ubiquitous as the tokamak, stellarator coils must overcome these issues.
 
The access properties of the set of coils which generate a prescribed magnetic surface refers to how that coil set restricts access to internal device components such as the first wall, divertors, blankets, and shields. Here, we define a coil set which has good access properties - that is, a coil set which allows for ease of access to the internal components of the device for maintenance, as being "open-access". Open-access coil sets  allow for ease of assembly of a device and studies of different designs of internal components.

The focus of this work is to explore the maximal limit of accessibility for a given equilibrium. The degree of accessibility is formulated as the fractional area of the winding surface populated by surface currents. A given equilibrium is found to have good access properties if the fraction of the winding surface needed to produce the equilibrium is small. The method is a combination of singular value decomposition (SVD) to ensure the surface currents used are low in amplitude and are placed in the most efficient locations on the winding surface to support the desired equilibrium and solution sparsification to find regions of the winding surface where surface currents may be removed from the winding surface. An example calculation using the HSX equilibrium is shown, finding that the equilibrium can be supported by placing currents on only 22\% of the HSX winding surface, assuming an external planar coil set provides the necessary toroidal field.

\subsection{Conditions on the coils}
Magnetic field coils must satisfy three properties: they must (i) produce magnetic surfaces (ii) be feasible in an engineering sense and (iii) be separated far enough from the plasma. In a fusion power plant, (iii) is a more difficult constraint than in existing devices. The coils must be sufficiently separated from the plasma to allow room for tritium breeding, neutron shielding, and device maintenance.

Magnetic confinement uses toroidal magnetic surfaces, also known as plasma surfaces. Toroidal magnetic surfaces are generated by magnetic field lines which wind around the torus without leaving the specified surface, $\mb{B}\cdot\h{n}=0$ where $\hat{n}$ is the unit normal to the desired magnetic surface. The magnetic field has contributions both from external currents and the plasma current.

In both tokamaks and stellarators the magnetic surfaces are tori, but in stellarators the tori are far more complicated with large variations in shape in the toroidal direction.

There is a practical limit on the distance between coils and the magnetic surfaces they produce. The magnetic field produced in a volume which contains no magnetic sources is curl-free, $\mb{B}=\nabla\Phi$, which when combined with $\nabla\cdot\mb{B}=0$ yields solutions for Laplace's equation in a cylinder. Remarkably, current potential solutions in cylindrical geometry do not vary significantly from results achieved in toroidal geometry \cite{EFDs:2016}. These solutions represent the current required to produce a given magnetic surface and increase exponentially with distance from the magnetic surface and in power of their Fourier harmonic. The further away currents are placed to produce an equilibrium, the less feasible they become \cite{EFDs:2016}\cite{Boozer:Stell_Design}. 

A feasible coil set cannot have (i) current densities that are beyond what is producible by coils (ii) forces on them that are too large (iii) curvatures or torsions that are too great and (iv) must be sufficiently widely separated.

It is important to have good chamber access for ease of assembly and maintenance. For example, the small plasma-chamber distance of W7-X was a contributor to budget overruns and schedule delays \cite{ZhuSimplerCoils}. Designing coils that optimize access is a focus of this work. Access is quantified by the fraction of the gridded winding surface used by shaping currents of current potential patches - the less of the winding surface used, the greater the access properties.

\section{Background}
The initial configuration space for 3D coils to produce a given magnetic surface is too large to search by brute force. Good coil guesses can be found using physics motivations. A typical method of doing this was invented by Rehker and Wobig \cite{ModCoils_1st:1973} in which a current sheet near the desired magnetic surface is found.

The formalism of these magnetic-surface-generating current sheets was developed by Merkel \cite{NESCOIL:1987} and used to successfully develop coil sets for W7-AS and HSX. Merkel extended the winding surface from infinitesimally separated, as in the Rehker and Wobig formulation, to a surface placed an arbitrary distance away from the desired magnetic surface. Merkel posed the problem as solving for what is now known as the current potential $\Phi$. Details of the formulation may be found In Landreman 2017 \cite{REGCOIL:2017}. 

$\Phi$ is known as the current potential, the magnetic surface generating stream function. 
\begin{equation}
\Phi(\theta,\zeta) = I_\theta\zeta/2\pi + I_\zeta\theta/2\pi + \Phi_{SV}(\theta,\zeta)
\label{eqn:Current_potential}
\end{equation}
and has two terms: the non-secular and secular (also known as single-valued and multi-valued) components. The secular components increase by $I_\theta$ for each toroidal transit and $I_\zeta$ for each poloidal transit of the torus. The non-secular components 
\begin{equation}
\Phi_{SV}(\theta,\zeta) = \sum_{m,n} \Phi_{m,n}^{c} \cos(m\theta-n\zeta) + \Phi_{m,n}^{s} \sin(m\theta-n\zeta)
\label{eqn:Current_potential_SV}
\end{equation}
are single-valued functions which are typically written as Fourier series. The non-secular terms are degrees of freedom for generating the desired magnetic field. The secular terms are degrees of freedom for the current potential. $I_\theta$ is fixed by the toroidal flux passing through the plasma. $I_\zeta$ is a degree of freedom to control whether the current potential is helical or not, and provides the magnetic flux passing through the hole in the torus. If $I_\theta=0$ the current potential is unable to produce toroidal flux. If $I_\theta,I_\zeta=0$ the current potential is single-valued.

The current density of a current potential is given as
\begin{equation}
\mb{K} = \nabla\Phi\times\h{n}
\label{eqn:CP_K}
\end{equation}
where $\h{n}$ is the unit normal vector to the winding surface. Large gradients in the current potential correspond to high current densities in the current sheet.


A prescribed magnetic surface is generated with currents by eliminating normal components of the magnetic field $\mb{B}$, which yields the normal field $\mb{B}_n = \mb{B}_{\mbox{toroidal}} \cdot \h{n}$ arising from the interaction between the toroidal field and magnetic surface geometry. In the simple situation that is studied here, the normal field arises from the axisymmetric toroidal field intercepting a non-axisymmertic magnetic surface. As a result, we focus on the single-valued current potential, which can produce the shaping of a magnetic surface when the toroidal magnetic field is pre-supplied. 

For the purposes of this work we restrict ourselves to only considering sources of $\mb{B}\cdot\h{n}$ from vacuum magnetic fields $\mb{B}=\nabla\Phi$ which is a purely toroidal $1/R$ field. A discussion on the differences in $B_n$ when considering other sources of toroidal field may be found in Sec. \ref{sec:Discussion}.  

Single-valued current potentials have equivalent representations as distributions of magnetic dipoles. This was shown, but not appreciated, in equation (7) of Merkel's 1987 paper \cite{NESCOIL:1987} and has been derived more rigorously in Zhu's 2020 work \cite{Zhu_PMs:2020}. Further expansions on the idea may be found \cite{EFDs:2016}, \cite{Boozer:Stell_Design}.

Given the field of a magnetic dipole,
\begin{equation}
\mb{B}(\mb{x}) = \dfrac{\mu_0}{4\pi} \dfrac{ 3(\mb{x}-\mb{x}') ((\mb{x}-\mb{x}') \cdot \mb{m}) - \mb{m} }{|\mb{x}-\mb{x}'|^3}
\notag
\end{equation}
an inductance matrix $\mb{L}_{\mbox{dip}}$ between magnetic dipoles placed on a winding surface $\mb{S}_w$ at points $\mb{x}_j$ and the field they create on a plasma surface $\mb{S}_p$ may be computed
\begin{equation}
\hspace{-0.75cm}
\mb{L}_{\mbox{dip}} = 
\begin{bmatrix}
\dfrac{3\h{\mb{r}}_{11}(\mathbf{m}_1\cdot\h{\mb{r}}_{11}-\mathbf{m}_1)}{|\mb{r}_{11}|^3} & \dfrac{3\h{\mb{r}}_{21}(\mathbf{m}_2\cdot\h{\mb{r}}_{21}-\mathbf{m}_2)}{|\mb{r}_{21}|^3} & ... \notag \\
\dfrac{3\h{\mb{r}}_{12}(\mathbf{m}_1\cdot\h{\mb{r}}_{12}-\mathbf{m}_1)}{|\mb{r}_{12}|^3} & ... & ... \notag \\
... & ... & ... \notag\\
\end{bmatrix} 
\notag
\end{equation}
where $\mb{r}=\mb{x}_j-\mb{S}_{p,i}$, $\h{\mb{r}}=\mb{r}/|\mb{r}|$, and $\mb{m}_j$ is the $j$th magnetic dipole vector. The inductance matrix is an $N\times M$ matrix where $N$ is the number of evaluation points on the magnetic surface $\mb{S}_p$ and $M$ is the number of dipoles used on winding surface $\mb{S}_w$. The winding surface may be represented as a mesh grid separated uniformly in $\theta$ and $\zeta$. The magnetic dipole winding surface then consists of $n_\theta$ by $n_\zeta$ points, each of which is populated by a single magnetic dipole. 

The problem then becomes a least-squares problem to minimize 
\begin{eqnarray}
\min\bigg( \big| \mb{L}_{\mbox{dip}}\cdot\mb{x} - \mb{B}_n \big|^2 \bigg)
\label{eqn:Dipoles_LSS}
\end{eqnarray}
where the components of $\mb{x}$ are the strengths of the dipoles on the grid.

Inductance matrices relate currents to the field they produce. For current potentials, $\mb{L}$ relates the current on the winding surface to the magnetic field produced on the surface $\mb{S}_p$. $\mb{L}$ relates the geometry of the two surfaces involved, the plasma surface $\mb{S}_p$ and the winding surface $\mb{S}_w$, and the gradient of the current potential $\nabla\Phi$. The inductance matrix can be very ill-conditioned and so represents the main computational difficulty of this least-squares problem. Further insights and a rapid way to compute the inductance matrix may be found here \cite{Cerfon_Multipole:2012}. 


For context, we note that the use of magnetic dipoles, specifically permanent magnets, to generate shaping fields is recently adopted approach to stellarator design \cite{Helander_PMs:2020}. In this approach, a surface or volume grid of magnetic dipoles is used to generate a given magnetic surface by varying dipole strength and orientation. These magnetic dipole methods have been improved and implemented for optimization \cite{Zhu_PMs:2020} \cite{Kaptanoglu_GPMO:2023} \cite{Hammond_PMs:2020} leading to the construction of the first permanent-magnet stellarator MUSE \cite{MUSE:2021}. Permanent magnet stellarators are remarkably able to produce magnetic surfaces to great accuracy and provide some open access, though are not viable for fusion power plants due to degradation of the magnets from neutron flux and the low field strength of the magnets if they are placed behind shielding. 

\section{Method}

\begin{figure}
	\includegraphics[width=2.75in]{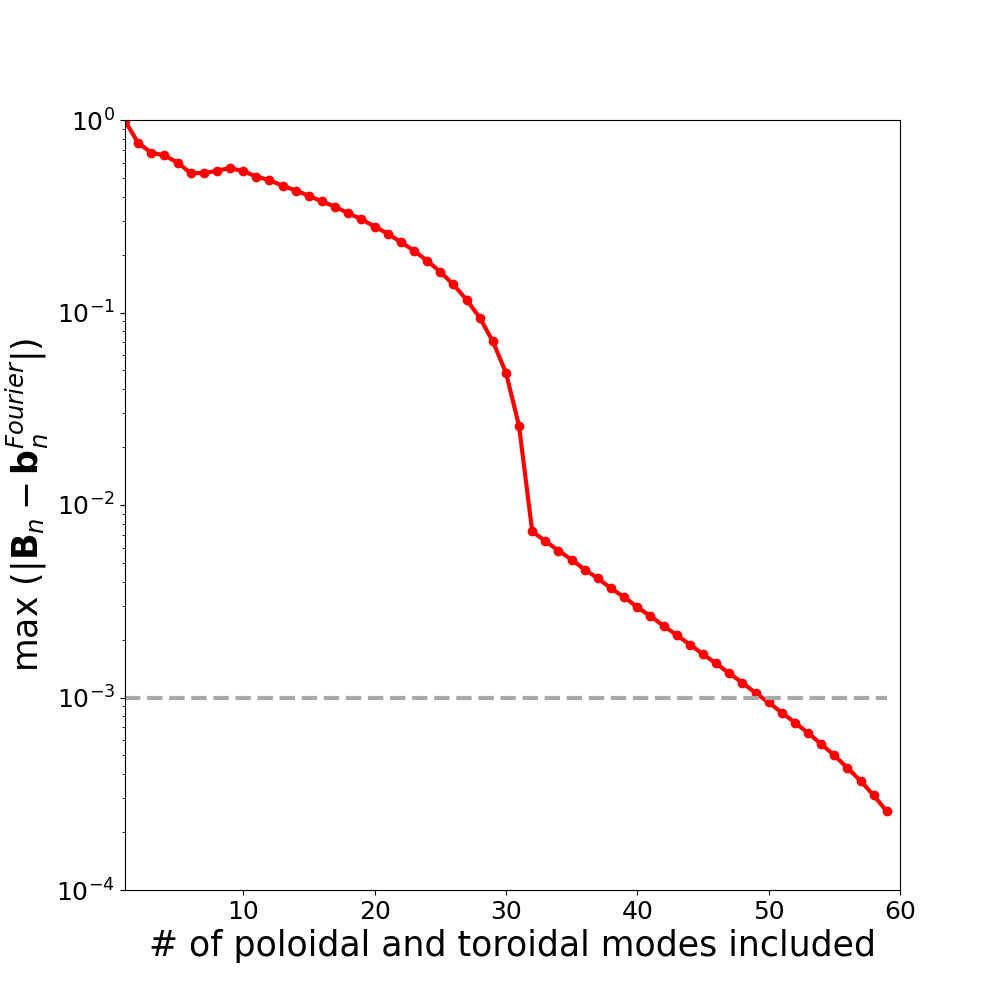}
	\caption{Fourier decomposition of Bn on the magnetic surface due to a toroidal flux. Achieving a maximum error of 0.1\% field accuracy is typically sufficient to reproduce the desired equilibrium. The HSX equilibrium used has some high mode number harmonics as can be seen in the not strictly logarithmic dependence of maximum $B_n$ defect with increasing Fourier resolution. This likely adds some complexity in the current potential patch solution.}
	\label{fig:Bn_Fourier}
\end{figure}

Magnetic dipoles are placed in a grid on, and oriented normal to, the HSX winding surface. 
Here, the HSX winding surface is the surface on which HSX's modular coils lie. A winding surface further back would produce more complicated current potentials while a closer surface produces simpler current potentials \cite{EFDs:2016}. A truncated version of HSX's experimental equilibrium was used as the original optimized version is lost, and current equilibria have built-in toroidal ripple. This truncated equilibrium minimizes coil ripple by removing Fourier components with more than 12 toroidal periods per field period. Despite this, as shown in Fig. \ref{fig:Bn_Fourier}, some ripple defects remain.

Stellarator symmetry may be used to account for each dipole's contribution from being mapped toroidally. As a result, only a half-field-period winding surface with a half-field-period of plasma surface need be considered when solving for the magnetic dipole strengths. The use of stellarator symmetry restricts the applicability of this inductance matrix formulation to only stellarator-symmetric equilibria, though the formulation is easily generalized to field-period symmetry.


The method described in the following subsections is closely related to the efficient field distributions method of \cite{EFDs:2016}, though this implementation uses the unique ability of current potential patches to construct discontinuous current sheets.

\subsection{Relation of individual dipole fields to $B_n$ patterns produced on the plasma surface}

The normal field $B_n$ on the plasma surface is Fourier decomposed. This step reduces computational complexity and is used to control the cancellation of the normal field errors to a tolerable level, typically a maximum $B_n/\avg{|\mb{B}_{\mbox{surf}}|}$ value of 0.1\%, where $\avg{|\mb{B}_{\mbox{surf}}|}$ is the average magnetic field strength on the magnetic surface. For this work, $\avg{|\mb{B}_{\mbox{surf}}|}=1$T.

Given a normal magnetic field $B_n(\mb{x})$, we Fourier decompose it as $B_n(\mb{x})=b_k(\theta,\zeta)$ where
\begin{equation}
b_k(\theta,\zeta) = \sum_{m,n}^{k}b_{m,n}\cos(m\theta-n\zeta)
\label{eqn:Bn_Fourier}
\end{equation}
Exponential convergence of the Fourier series allows us to choose a Fourier truncation value $k$ such that the maximal error $\delta$ between the normal field and its Fourier representation is $|B_n - b_k|^2 < \delta$. 

Next, the inductance matrix  $\mb{L}_{\mbox{dip}}$ is similarly Fourier decomposed to express the relationship between individual magnetic dipoles and the $B_n$ patterns they cast globally on the plasma surface. This matrix is referred to as the Fourier-decomposed inductance matrix $\mb{L}$.

These steps of re-expressing the normal field on the surface and inductance matrix in terms of Fourier series allows for the identification of efficient patterns of magnetic dipole placement by relating the strength of each dipole to their effect on the plasma surface globally. By efficiency, we mean that dipoles placed at these locations on the winding surface produce the largest amount of $\mb{B}_n$ reduction on the plasma surface.

\subsection{Singular value decomposition truncation}

\begin{figure}
	\includegraphics[width=2.75in]{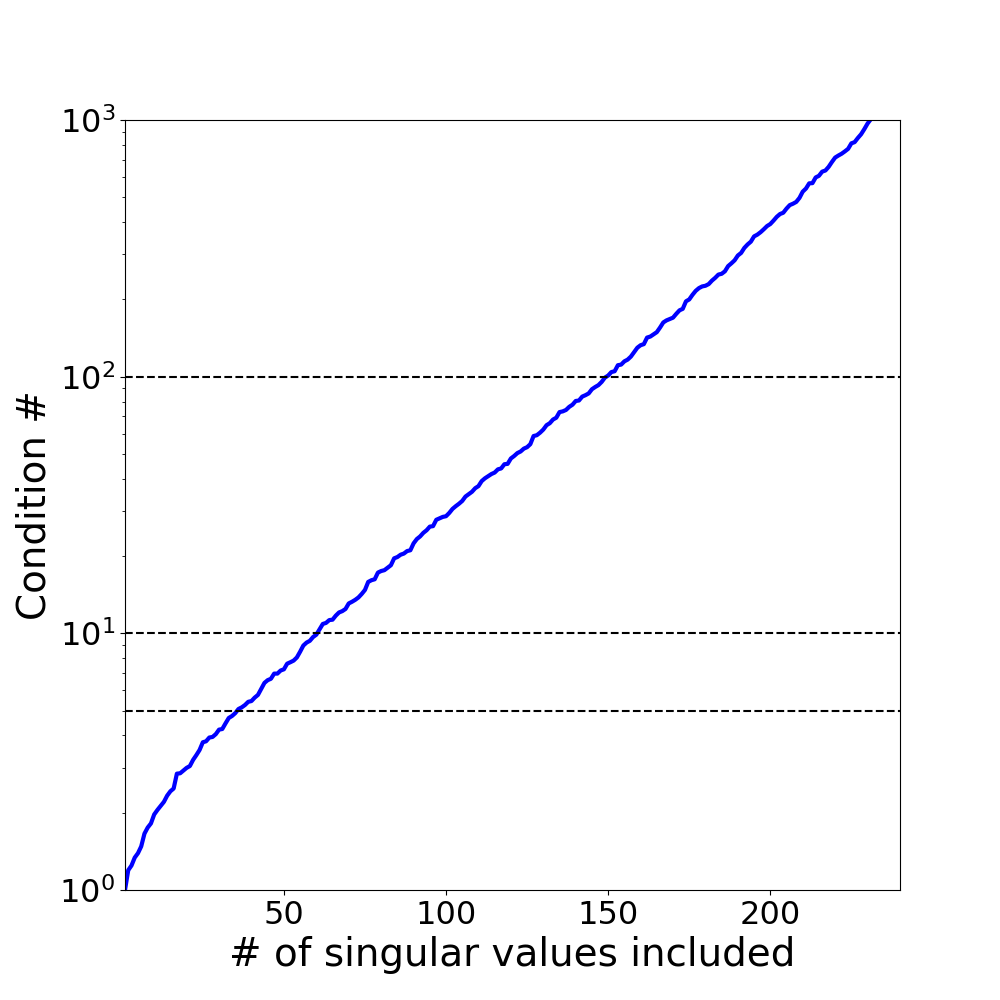}
	\caption{Condition number of the Fourier-decomposed inductance matrix. Including more singular vectors results in an inductance matrix with a higher condition number, though in exchange $B_n$ decreases and the dipole patterns increase in complexity}
	\label{fig:CondNumber}
\end{figure}

In what follows we describe the the singular value decomposition (SVD) truncation of the Fourier decomposed inductance matrix. 

The inductance matrix is decomposed using singular value decomposition 
\begin{equation}
\mb{L} = \mb{U}\mathbf{\Sigma}\mathbf{V}^*
\label{eqn:SVD_formula}
\end{equation}
From this decomposition, the singular vectors corresponding to the largest singular values are locations of efficient magnetic dipole placement on the winding surface to support the magnetic surface. 

A new truncated inductance matrix is then constructed using only $n$ values of the SVD-truncated inductance matrix, 
\begin{equation}
\mb{L}_{\mbox{trunc}} = \sum_{i=1}^{n} \sigma_i \mb{u}_i \mb{v}_i^*
\label{eqn:SVD_formula_truncated}
\end{equation}
where $n$ is chosen according to the desired condition number of the inductance matrix, Fig. \ref{fig:CondNumber}. A higher condition number results in the inclusion of more rapidly-varying patterns of magnetic dipole amplitudes in the winding surface, corresponding to higher current densities. As such, it is desirable to keep the condition number low, such as between 5-25. 

With a given condition number chosen,  $\mb{L}_{\mbox{trunc}}$ from equation \ref{eqn:SVD_formula_truncated} is then used to solve for the amplitudes of each magnetic dipole,
\begin{equation}
\min\bigg( | \mb{L}_{\mbox{trunc}}\cdot\mb{x}-\mb{b}_k |^2 \bigg)
\label{eqn:EFD_dipole_LSS}
\end{equation}

\begin{figure*}[t]
	\hspace*{-1cm}
	\includegraphics[width=\textwidth]{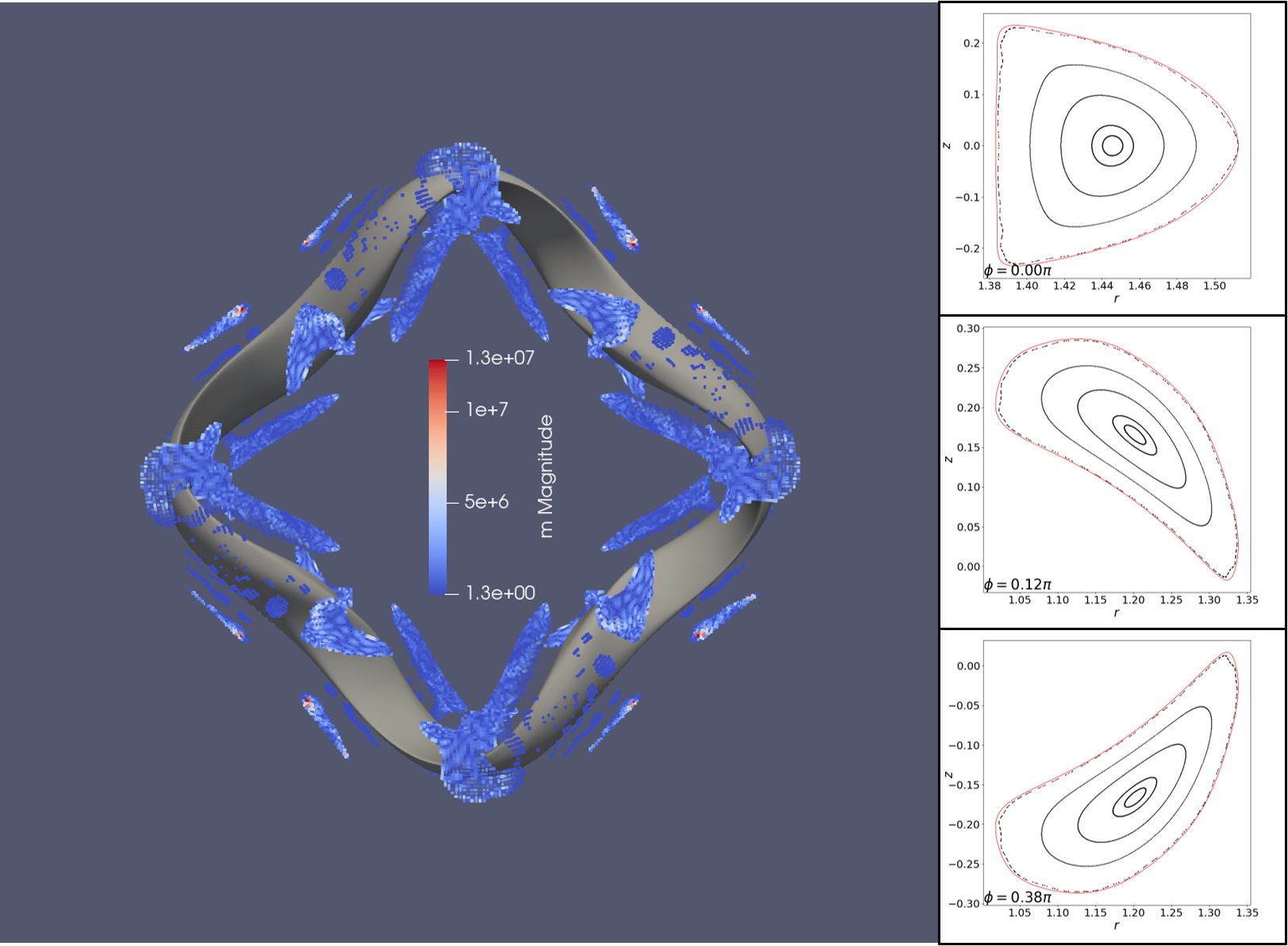}
	\caption{Configuration with current potential patches covering only 22\% of the winding surface. The dipole strength is shown on the color bar and may be related to flowing currents through equation \ref{eqn:CP_K}. The magnetic surface, shown here as the magnetic surface, is reproduced to a maximum defect of $B_n/\avg{|\mb{B}_{\mbox{surf}}|}=0.1\%$. Poincare plots are shown for three toroidal angles along the right side of the figure, with the actual HSX equilibrium outlined in red. The Poincare plots were generated by field line tracing a $1/R$ toroidal magnetic field with the shaping fields generated by current potential patches.} 
	\label{fig:Dips_3D_combined}
\end{figure*}

\subsection{Solution sparsification}
Here we describe the dipole removal method. To begin, a threshold dipole strength $|\mb{m}_{\mbox{thresh.}}|$ is chosen.  Each dipole amplitude $|x_i|$ of $\mb{x}$ resulting from minimizing equation \ref{eqn:EFD_dipole_LSS} is then compared to the threshold dipole strength. Dipole strengths less than the threshold, $|x_i|<|\mb{m}_{\mbox{thresh.}}|$, are then zeroed out, $|x_i|=0$, enforcing solution sparsity. In effect, this 'removes' the dipole from the winding surface. Then, a new least-squares problem is solved for the remaining dipole amplitudes $\mb{x}_{\mbox{fraction}}$ for a final adjustment calculation after solution sparsification has been implemented. 

Choosing the dipole strength threshold $|\mb{m}_{\mbox{thresh.}}|$ is a tradeoff between solution sparsity, maximal dipole strengths, and solution accuracy. In the example shown, it is chosen to correspond to a low current density current sheet with 22\% coverage and a 0.1\% field error. The current density of the patch is related by the gradient of dipole field strengths. Generally, forcing low dipole strengths produces solutions which cover larger surface areas than solutions which allow high dipole strengths.

\section{Results \label{sec:SVD_fractional_regularization}}

\begin{figure}
	\includegraphics[width=3.25in]{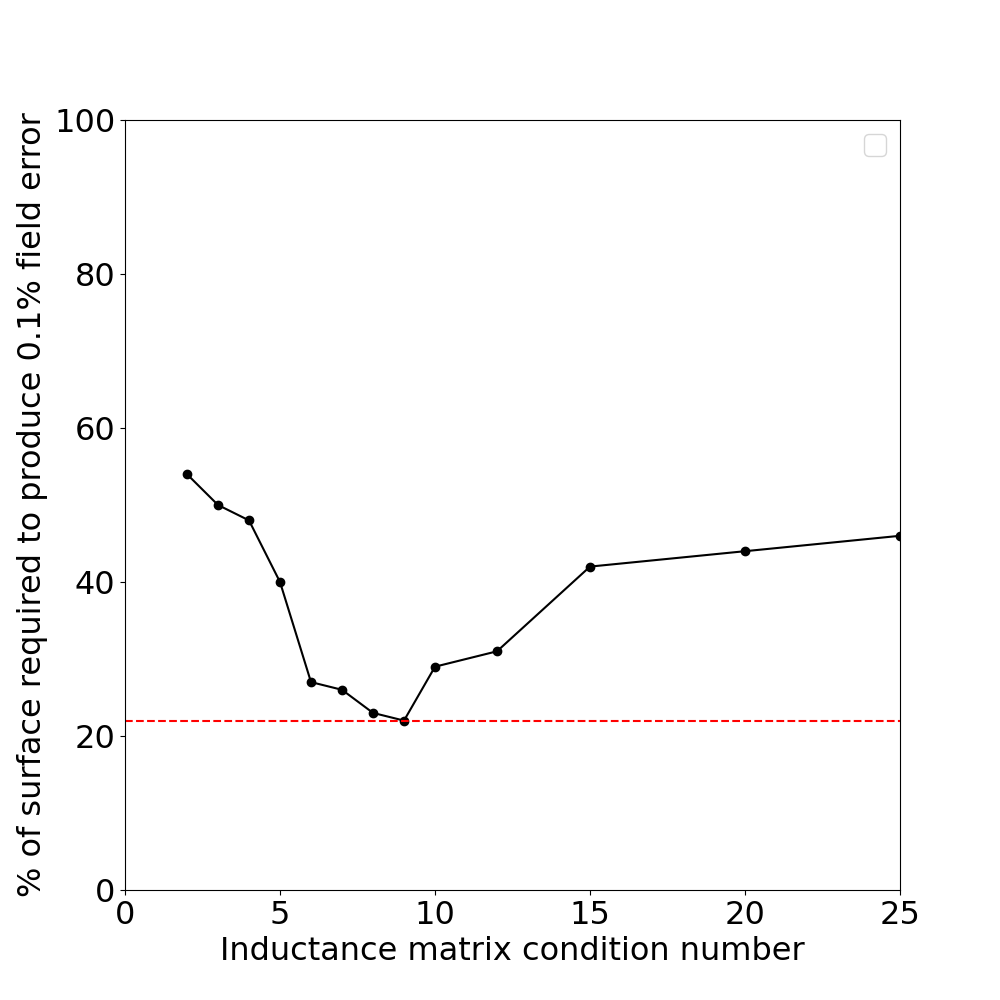}
	\caption{Percentage of winding surface coverage compared to the condition number of the inductance matrix used. For condition numbers less than 6, an insufficient number of efficient modes of the inductance matrix are used to reduce $B_n$ sufficiently. For condition numbers 10 and above, dipoles are placed inefficiently. A minimum coverage of 22\% for an inductance matrix with condition number 6 is found as the best achievable result.}
	\label{fig:Frac_vs_CondNumb}
\end{figure}

A calculation is first made covering the whole surface with dipoles and it is found that a maximum error of  0.1\% is achievable at a condition number of 9.  The dipoles with the smallest magnitudes are removed and the remaining dipole amplitudes are re-calculated, resulting in Fig. \ref{fig:Dips_3D_combined}.  The fraction that are retained gives the fraction of the wall area that is covered by dipoles, finding that only 22\% winding surface coverage is necessary.

To maintain a maximum error of 0.1\% a certain fraction of the surface must be covered as a function of condition number.  As shown in Fig. \ref{fig:Frac_vs_CondNumb}, this fraction reaches a minimum at approximately 22\% coverage and actually increases if the condition number is made larger. Using inductance matrices with higher condition numbers yields less-sparse solutions as less-efficient dipole patterns are driven which necessarily increases the amplitudes of amplitudes. In turn, when thresholding the dipole amplitudes to enforce sparsity, high-strength dipoles are placed in inefficient locations for minimizing $B_n$. Conversely, lower condition number inductance matrices than 10 do not contain enough variation to minimize $B_n$ to a tolerable level. 

The ideal inductance matrix condition number for maximum port access depends on both the magnetic surface being targeted and the winding surface on which the dipoles are placed. Thus, a figure such as Fig. \ref{fig:Frac_vs_CondNumb} should be computed to identify the ideal condition number. 


Per half-field-period, the majority of currents are placed on only one the top, or bottom, of the winding surface depending on which half-period of the winding surface is considered. For example, for the first half-field-period, most of the current potential patches lie on the top of the winding surface. They accordingly provide most shaping for both the top and bottom of the magnetic surface from this upper position on the winding surface. The underside is then largely free port space. For the second half-field-period, the same story holds but with the top and bottom reversed. 

The density of magnetic dipoles on the surface $n_d$ influences the error field significantly if two few dipoles are used. The interpretation of this effect is that higher dipole densities can support shorter-wavelength Fourier spectra.  As we populate the winding surface with more and more dipoles ($n_d\to\infty$), the discrete set of dipoles approximates a continuous finite-element-method basis, though the improvements in $B_n$ reduction do saturate.

\section{Discussion \label{sec:Discussion}}
A novel method to investigate access properties of stellarator coil sets was developed. The tool, known as current potential patches, uniquely retain the solution properties of current potentials for coil design yet are able to generate sparse solutions. An example calculation using the HSX winding and magnetic surfaces demonstrates the ability of current potential patches to (i) identify crucial locations of current placement for surface shaping, for example using windowpane coils and (ii) provide an example bounding calculation on a coil set with maximal open-access properties which still produces the desired magnetic surface to a given degree of accuracy. Overall, this method has use in magnetic confinement fusion device design and specifically aids in designs with access to the vacuum vessel and region interior to the coil set. 

The access limits set by the method are approximate due to the use of surface currents. The use of filamentary coils would likely decrease access properties due to the inability of filamentary coils to produce the current distributions of current sheets. 

A source $1/R$ toroidal field was used here to generate $B_n$ on the plasma surface. Using toroidal field coils in a realistic coil set would slightly change the $B_n$ field by introducing toroidal ripple. Other work, not shown here, using helical coils and a helical current potential showed only small differences in the $B_n$ pattern and so were not included.

Results could be improved further by using some degrees of freedom in the toroidal field coil set, though this is not the focus of this work. For example, toroidal field coils can remain planar but be shaped and tilted to account for some of the field shaping. Further, a minimal number of toroidal field coils could be found in conjunction with current potential patches minimizing toroidal field ripple to promote good access properties.

A surprising result of this work is that low current density features in current potentials are associated with efficient placement of the coil, and conversely that high current density features inefficiently reduce the global $B_n$. This is not surprising in the context of efficient fields \cite{EFDs:2016}, though understanding its implication for filamentary coil design is important. For example, the coil design of Thea Energy \cite{Thea} uses grids of windowpane coils, analogous to a current sheet. This work suggests that when designing port space for such a design, it is more prudent to remove high-current windowpane coils than low-current coils.

This method has many applications, a few of which we describe here. For one, coil sets consisting of simple toroidal field coils or helical coils in conjunction with windowpane coils can be generated using this method. In this case, contours of $\Phi_{SV}$ are the contours of windowpane coils. Second, a metric describing the amenability of an equilibrium to open-access coil sets could be derived. Third, this method can be used to identify locations on the winding surface which may be replaced by ports.  Fourth, though not the scope of this paper, alternative ways to generate shaping currents, such as using flowing liquid metal analogous to surface currents \cite{Kolemen:LMCurrents}, directly benefit from this method and provide a measure of maximal port space for such configurations. Fifth, this method can be applied to error field and toroidal ripple minimization in tokamaks. For example, the error field of the ITER  toroidal field coil set could be measured and effectively mitigated using windowpane coils generated from current potential patches, or to minimize the toroidal field ripple between adjacent toroidal field coils.


\section*{Acknowledgments}
This material is based upon work supported by the U.S. Department of Energy, Office of Science, Office of Fusion Energy Sciences under Awards DE-FG02-95ER54333 and DE-FG02-03ER54696. \\






\section*{Data availability statement}

Data available on request from the authors.


\end{document}